\documentclass[usenatbib]{mn2e}
\usepackage{graphicx}
\usepackage{amsmath}
\usepackage{tabularx}
\usepackage{float}
\usepackage{url}
\newcolumntype{L}{>{$}l<{$}}
\newcolumntype{R}{>{$}r<{$}}

\def\apj{\rm ApJ}

\def\aj{\rm AJ}
\def\mnras{\rm MNRAS}

\def\pasp{\rm PASP}

\def\aap{\rm AAP}
\def\aaps{\rm A\&AS}

\def\gax{\mathrel{\raise.3ex\hbox{$>$}\mkern-14mu\lower0.6ex\hbox{$\sim$}}}
\def\lax{\mathrel{\raise.3ex\hbox{$<$}\mkern-14mu\lower0.6ex\hbox{$\sim$}}}
\def\gtorder{\mathrel{\raise.3ex\hbox{$>$}\mkern-14mu
             \lower0.6ex\hbox{$\sim$}}}
\def\ltorder{\mathrel{\raise.3ex\hbox{$<$}\mkern-14mu
             \lower0.6ex\hbox{$\sim$}}}

\begin{document}

\title 
   [An All-Sky Search For R Coronae Borealis Stars in ASAS-SN]
   {An All-Sky Search For R Coronae Borealis Stars in ASAS-SN}

\author[J. V. Shields et al.]{J. V. Shields$^{1,2}$\thanks{E-mail: shields.217@osu.edu},
T. Jayasinghe$^{1,2}$,
K. Z. Stanek$^{1,2}$,
C. S. Kochanek$^{1,2}$,
B. J. Shappee$^{3}$,
\newauthor 
T. W. -S. Holoien$^{4}$,
Todd A. Thompson$^{1,2}$,
J. L. Prieto$^{5,6}$,
Subo Dong$^{7}$
\\
$^{1}$Department of Astronomy, The Ohio State University, 140 West 18th Avenue, Columbus, OH 43210, USA\\
$^{2}$Center for Cosmology and Astroparticle Physics, The Ohio State University, 191 W. Woodruff Avenue, Columbus, OH 43210, USA\\
$^{3}$Institute for Astronomy, University of Hawaii, 2680 Woodlawn Drive, Honolulu, HI 96822,USA\\
$^{4}$Carnegie Observatories, 813 Santa Barbara Street, Pasadena, CA 91101, USA\\
$^{5}$N\'ucleo de Astronom\'ia de la Facultad de Ingenier\'ia y Ciencias, Universidad Diego Portales, Av. Ej\'ercito 441, Santiago, Chile\\
$^{6}$Millennium Institute of Astrophysics, Santiago, Chile\\
$^{7}$Kavli Institute for Astronomy and Astrophysics, Peking University, Yi He Yuan Road 5, Hai Dian District, China\\
}

\maketitle

\begin{abstract}
We report the discovery of 19 new R Coronae Borealis (RCB) star and DY Per candidates with light curves from the All-Sky Automated Survey for Supernovae (ASAS-SN). We examined both an existing set of 1602 near/mid-IR selected candidates and an additional 2615 candidates selected to have near/mid-IR SEDs consistent with those of known R Coronae Borealis stars. We visually inspected the light curves for the characteristic variability of these systems.

\end{abstract}

\begin{keywords}
Catalogues - Stars: variable -AGB and post-AGB - carbon
\end{keywords}

\section{Introduction}
\label{sec:introduction}

The R Coronae Borealis (RCB) stars are a rare class of hydrogen deficient red giants. These stars are characterized by dramatic, unpredictable photometric declines with slow returns to full luminosity caused by clouds of carbon dust forming and flowing away from the star (e.g., \citealt{1997MNRAS.285..317F}, \citealt{2012JAVSO..40..539C}, \citealt{2007A&A...466L...1L}). These declines can be as great as 9 magnitudes in V band, and can last from a month to hundreds of days (\citealt{Tisserand2012}). 

There are two possible evolutionary paths for RCBs. RCBs could be the products of a final helium-flash in heavily evolved single stars before they cool to become white dwarfs. Alternatively, they could be merger products of lower mass He white dwarfs with higher mass CO white dwarfs. The abundance of \textsuperscript{18}O in cool RCBs heavily favors the latter theory (\citealt{2007ApJ...662.1220C}, \citealt{2010ApJ...714..144G}). 
Additionally, the He-rich pre-white dwarf KPD 0005+5106 has the abundances expected for a double degenerate merger. This both confirms that it is the descendant of an RCB star and reinforces the merger model for their origin (\citealt{2015A&A...583A.131W}). Observations of R Coronae Borealis, the prototype of its class, during a photometric minimum revealed a possible planetary nebula around the star (\citealt{2011ApJ...743...44C}). This is not predicted by the merger model and may support the He flash model. However, \cite{2015AJ....150...14M} found that the circumstellar shell is most likely not a fossil planetary nebula, but is instead a result of RCB phase mass loss. Thus, the double degenerate merger scenario is presently the favored explanation of RCB stars.

If the double degenerate scenario is correct, merger and lifetime arguments predict between 100 and 500 RCB stars in our galaxy (\citealt{2018arXiv180901743T}, \citealt{2018arXiv180711514L}, \citealt{2015ApJ...809..184K}). At present, there are 117 RCB stars known in the Galaxy and 30 in the Magellanic Clouds (\citealt{2018arXiv180901743T}). The number of known RCB stars has more than doubled in the past decade (\citealt{Tisserand2008}), and many of these new RCBs were found by searching for a combination of a mid-infrared excess and variability (\citealt{Tisserand2013}, \citealt{2016IBVS.6190....1N}, \citealt{2014JAVSO..42...13O}) using techniques developed in \cite{Tisserand2012}. Here we expand this approach to the full sky using ALLWISE (\citealt{2010AJ....140.1868W}) and 2MASS (\citealt{2006AJ....131.1163S}) to photometrically select candidates, and ASAS-SN (\citealt{2014ApJ...788...48S}, \citealt{2017PASP..129j4502K}) to examine their optical variability. Simultaneously with this work, \cite{2018arXiv180901474T} also recently used the complete WISE dataset to update their RCB candidates, and they report 45 spectral confirmations in \cite{2018arXiv180901743T}. Although light curves can be an excellent indicator, only a spectroscopic follow-up can confirm the identification of RCB stars.

The All-Sky Automated Survey for SuperNovae (ASAS-SN) is a ground based survey hosted by Las Cumbres Observatory (\citealt{2013PASP..125.1031B}) that has been monitoring the entire sky on a 2-3 day cadence to a depth of V $\leq$ 17 mag since 2013 using two units consisting of 4 telescopes in a common mount located in Hawaii and Chile. ASAS-SN has recently expanded, adding 3 more units located in Chile, Texas, and South Africa, respectively. ASAS-SN was created to monitor the sky for bright supernovae, but it also continuously monitors for variable stars (\citealt{2018MNRAS.tmp..817J}). 
In this work we search for new RCB stars. In Section 2 we outline the photometric selection of the candidates. In Section 3 we present our list of RCB candidates and their ASAS-SN light curves.

\section{Target Selection}
\label{Target Selection}

We started with the 2MASS and WISE selected list of 1602 candidates from \cite{Tisserand2012}. The selection method required that each source had data in all 7 bands (J, H, K, and W1-W4), and selected for stars with infrared properties similar to known RCBs, taking into account interstellar redenning by Galactic latitude. Cuts were made to reject other stars with similar infrared colors such as Asymptotic Giant Branch stars and Miras. 

\cite{Tisserand2012} selected these candidates before WISE data were available for the full sky. To select across the full sky, we used an alternate approach, simply looking for stars with
spectral energy distributions (SEDs) similar to those of known RCBs.  We started from the nominal list of ``known'' RCBs from 
SIMBAD (\citealt{2000A&AS..143....9W}), albeit with the knowledge that some of these classifications
were likely problematic, and objects from the ALLWISE (\citealt{2010AJ....140.1868W}) catalog with
defined WISE and 2MASS magnitudes satisfying a somewhat broader version
of \citealt{Tisserand2012}'s first color cut, namely,
\begin{equation}
\begin{split}
 W2-W3>0.75, \\ 
W2-W3 < 3.00, \\
 W3-W4 < 1.30, 
\end{split}
\end{equation}
and none of their other criterion.  This provided a list of 93 ``known'' RCBs and roughly 1.3 million WISE sources. 

The SED of a new RCB can
differ from that of a known RCB due to changes in luminosity, distance
and extinction, where the change in extinction can either be due to changes
in either Galactic or circumstellar extinction.  We do not differentiate
between the two sources of extinction since the differences are primarily due to
changes in the physics of scattered photons, which are less important in
the infrared (see the discussion in Kochanek et al. 2012).

We assume that changes in extinction only modify the 2MASS magnitudes as
a simplifying assumption.  So for each ALLWISE source and each ``known'' 
RCB, we first use the WISE magnitudes to estimate a change
in distance and luminosity as
\begin{equation}
    \Delta \mu = { \frac{1}{4} } \sum_{i=1,4} \biggl( W_i(WISE) - W_i(RCB_j)\biggr)
\end{equation}
which assumes uniform weighting of the four WISE bands ($W_i$) since this
exercise is almost certainly dominated by systematic errors. Then, with
$\Delta\mu$ fixed, we determine the change in extinction which would best
match the near-IR magnitudes ($M_i=J$, $H$, and $K_s$), 
\begin{equation}
\begin{split}
  \Delta E & = \biggl[ \sum_{i=1,3} R_i ( M_i(WISE) - M_i(RCB_j) - \Delta\mu )\biggr] \\
          &  \biggl[ \sum_{i=1,3} R_i^2 \biggr]^{-1}
\end{split}
\end{equation}
where the $R_i$ are the extinction coefficients.  We then computed the
root-mean-square magnitude residual $\sigma_j$ for each of the trial $j=1 ... 93$
RCBs corrected for the number of degrees
of freedom after fitting two parameters ($\Delta\mu$ and $\Delta E$).
We accepted an object as an RCB candidate if any $\sigma_j<0.2$~mag,
as this recovered 82 of the 93 ``known'' RCBs if we used this
method to search for each of them after excluding the star being tested
from the SED match.

For each of the ``known'' RCBs we then counted how many candidates were associated with it and iteratively eliminated stars producing 
too many candidates for new RCBs to be useful. As expected, we found that the SIMBAD listing is contaminated 
by sources other than RCBs.  For example, the worst comparison
star was MACHO118.18666.100, with 235 thousand (!) matches, 
which \cite{Tisserand2008} found to be an M giant.  In fact, all
of the ``known'' RCBs producing such large numbers of matches
are reported to be other sorts of variables (SY~Hyi as a semi-regular
variable, \citealt{Lawson1989}, V618~Sgr as a symbiotic star, \citealt{Kilkenny1997},
V1317~Sco as a Mira, \citealt{Tisserand2013}, V589~Sgr as a symbiotic
star, \citealt{Mennickent2001}, AE~Cir as a symbiotic star, \citealt{Mennickent2008}, 
GM~Ser as a Mira, \citealt{Tisserand2013}, and TYC6283-1417-1 as a Mira,
\citealt{Tisserand2013}).  With the last of these, the maximum number of
matches had dropped to 11 thousand. We also dropped LT~Dra, where the
origin of its classification is unclear and whose variability is reported
to be spurious by the AAVSO (\citealt{2002AAS...201.1711H}). 

Next there were ``known'' RCBs where we could find no arguments
that they were misclassified but which still produced too
many matches for a feasible search.  Many of these 
(in order of numbers of matches, Y~Mus, SV~Sge, XX~Cam, MACHO308.38099.66
and EROS2-CG-RCB-12) were also dropped by \cite{Tisserand2012} as falling outside
their color selection criteria.  We also dropped V409~Nor, HV5637, which was spectroscopically
confirmed as an RCB star by \cite{1972MNRAS.158P..11F}, OGLE BUL-SC 37 133492\footnote{First listed the RCB candidated OGLE-GC-RCB-Cand-1 by \cite{2011A&A...529A.118T}. It is still not a spectroscopically confirmed RCB star.}, EROS2-LMC-RCB-8, EROS2-CG-RCB-2,
V1405~Cyg, ASAS-RCB-18, and MACHO135.27132.51.  In total we rejected 24
of the initial list of 93 ``known'' RCBs.
This left us with a list of 2615 candidates, 65 of which are the remaining, ``known''
RCBs (which survive this process by its very definition). The
list also includes 609 of the color-selected candidates from \cite{Tisserand2012}.

\begin{figure*}
\centering
\includegraphics[width=\textwidth]{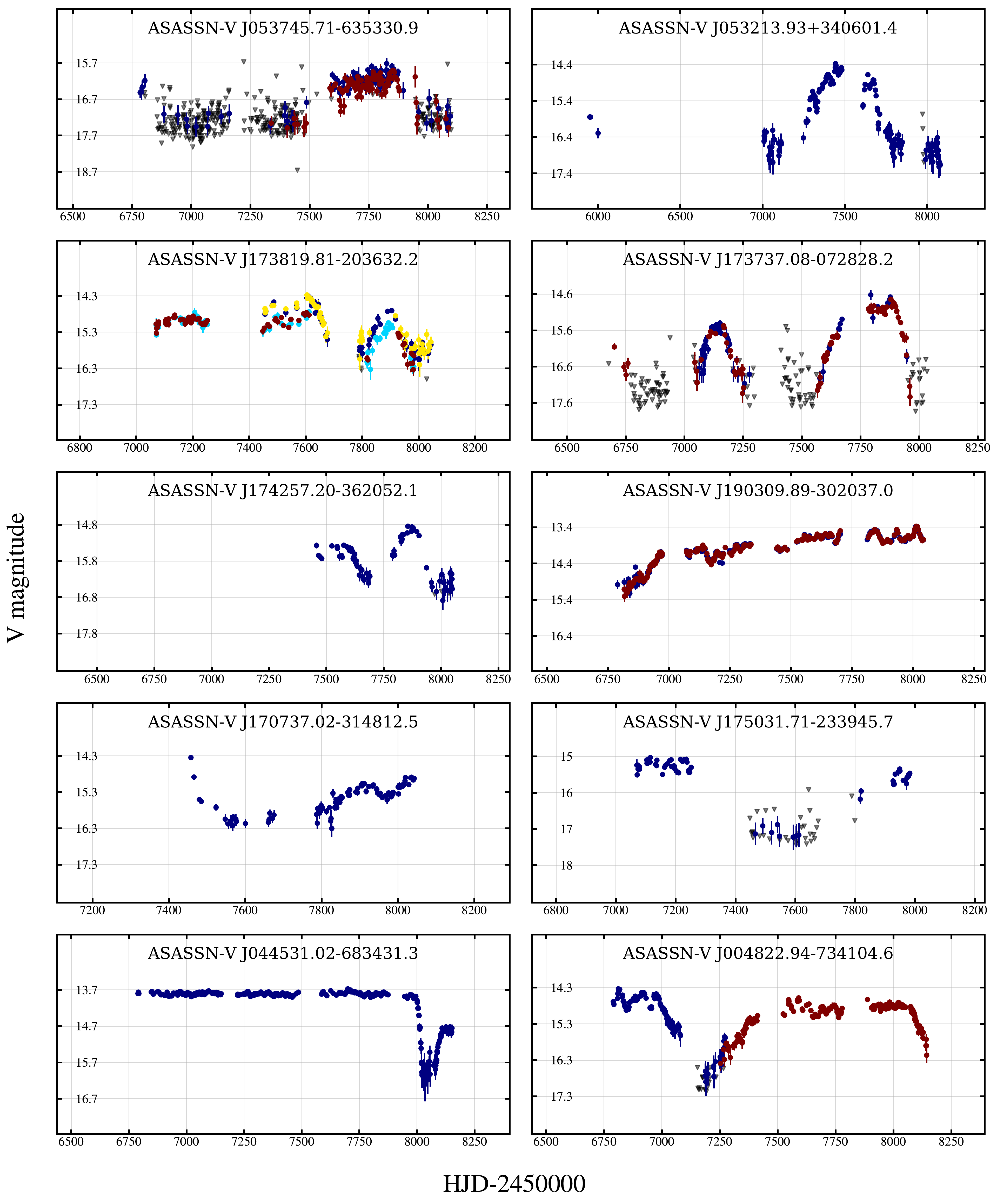}
\caption{ 
	ASAS-SN V-band light curves of the strong RCB candidates from Table \ref{table:rcbs}. All panels have the same dynamic range in magnitude. The different colors represent different ASAS-SN cameras. 
}
\label{fig:rcb1}
\end{figure*}

\clearpage

\begin{figure*}
\centering
\includegraphics[width=\textwidth]{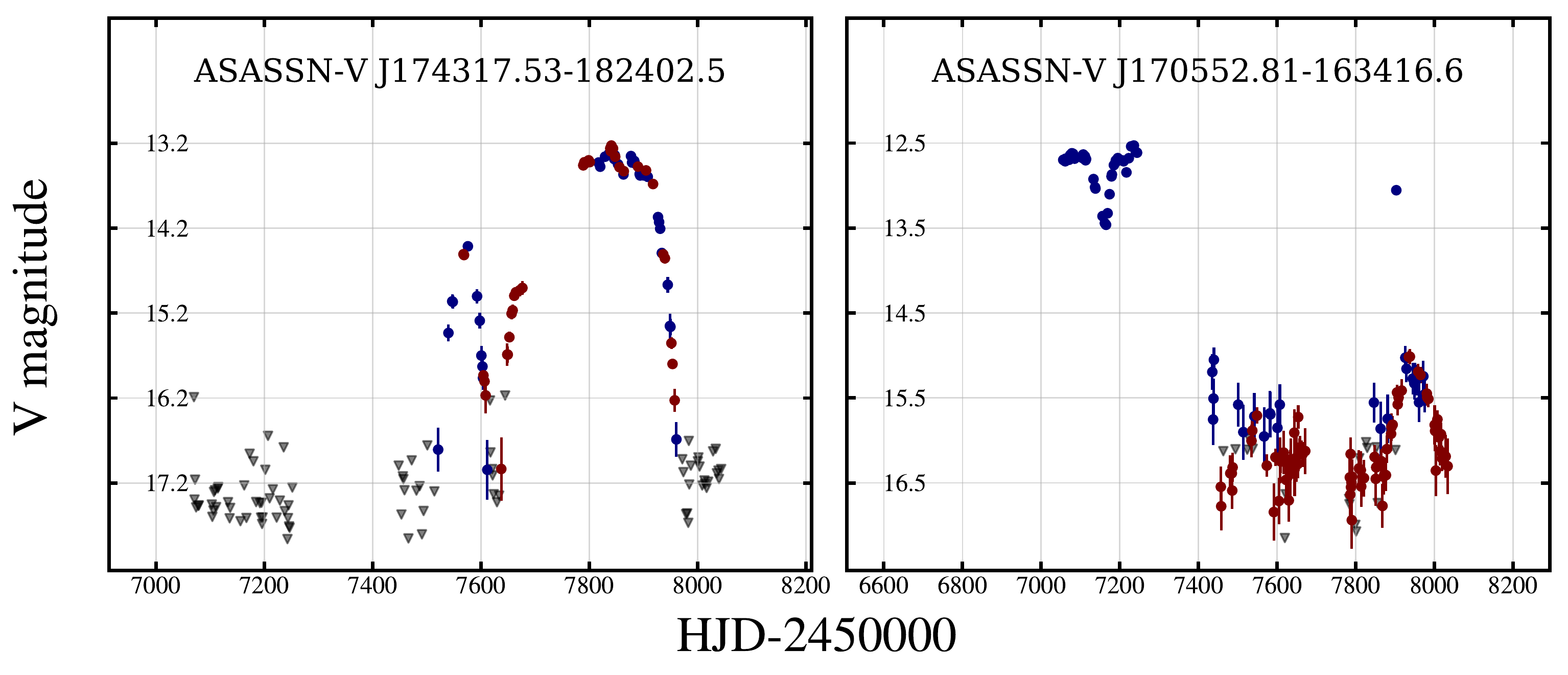}
\caption{ 
	ASAS-SN V-band light curves of the strong RCB candidates from Table \ref{table:rcbs} with large amplitude variation. These panels have a larger vertical scale than the other figures.
}
\label{fig:rcb_large}
\end{figure*}

\begin{figure*}
\centering
\includegraphics[width=\textwidth]{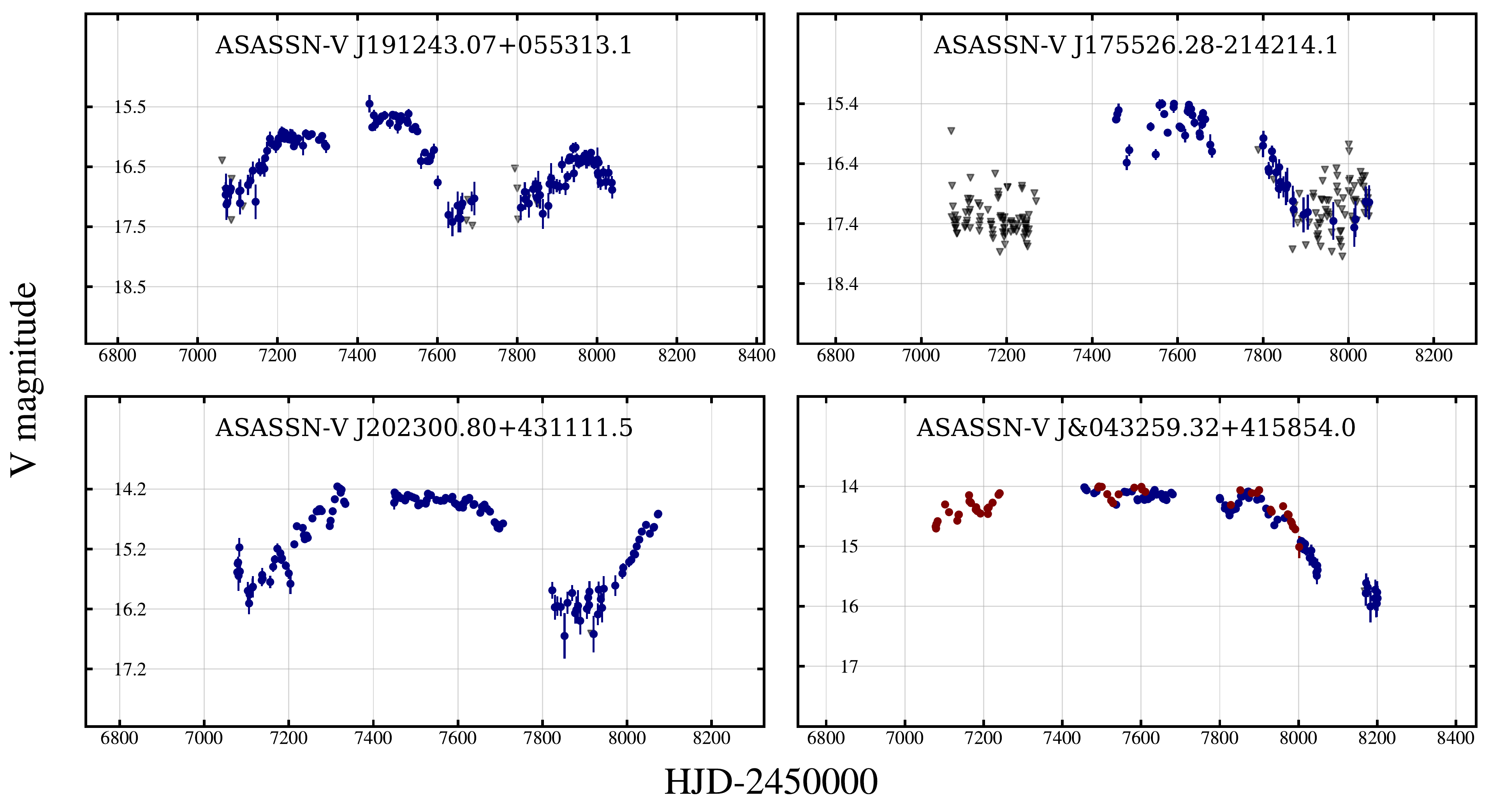}
\caption{ 
	ASAS-SN V-band light curves of the DY Per candidates from Tables \ref{table:rcbs} and \ref{table:old}. 
  }
\label{fig:dy_per}
\end{figure*}

\begin{figure*}
\centering
\includegraphics[width=\textwidth]{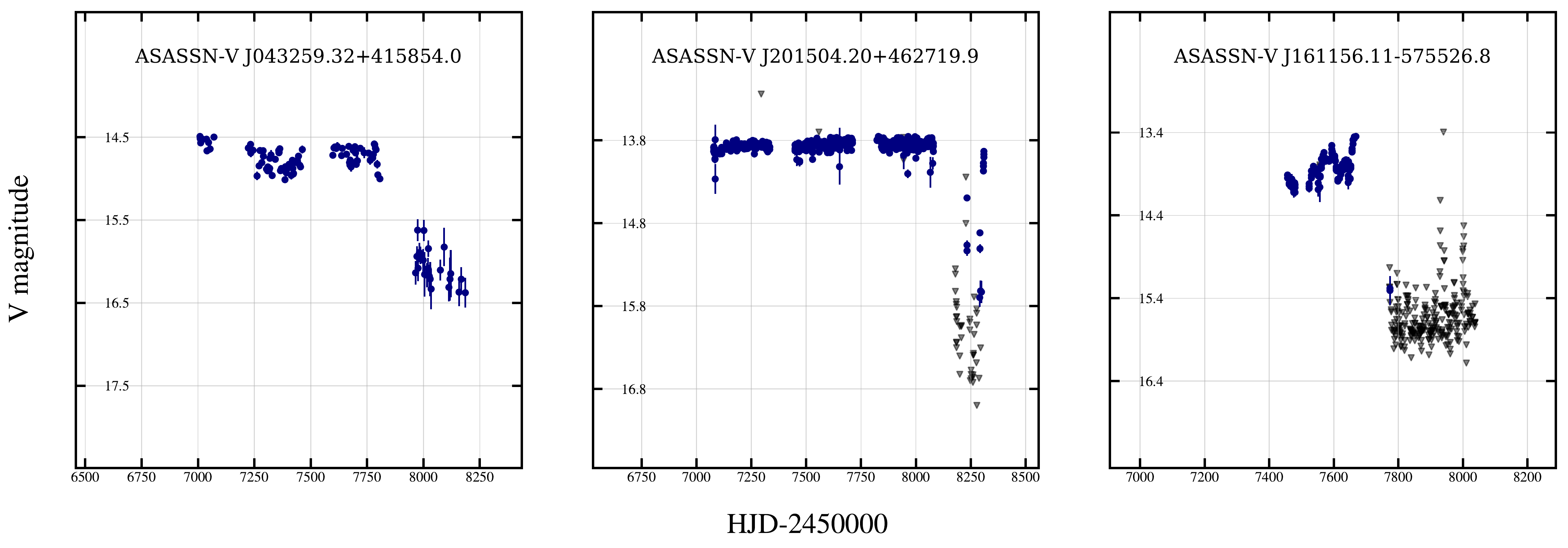}
\caption{
	ASAS-SN light curves of the RCB candidates discovered outside of the initial search from Table \ref{table:old}.
    }
\label{fig:rcb_old}
\end{figure*}

\begin{figure*}
\centering
\includegraphics[width=\textwidth]{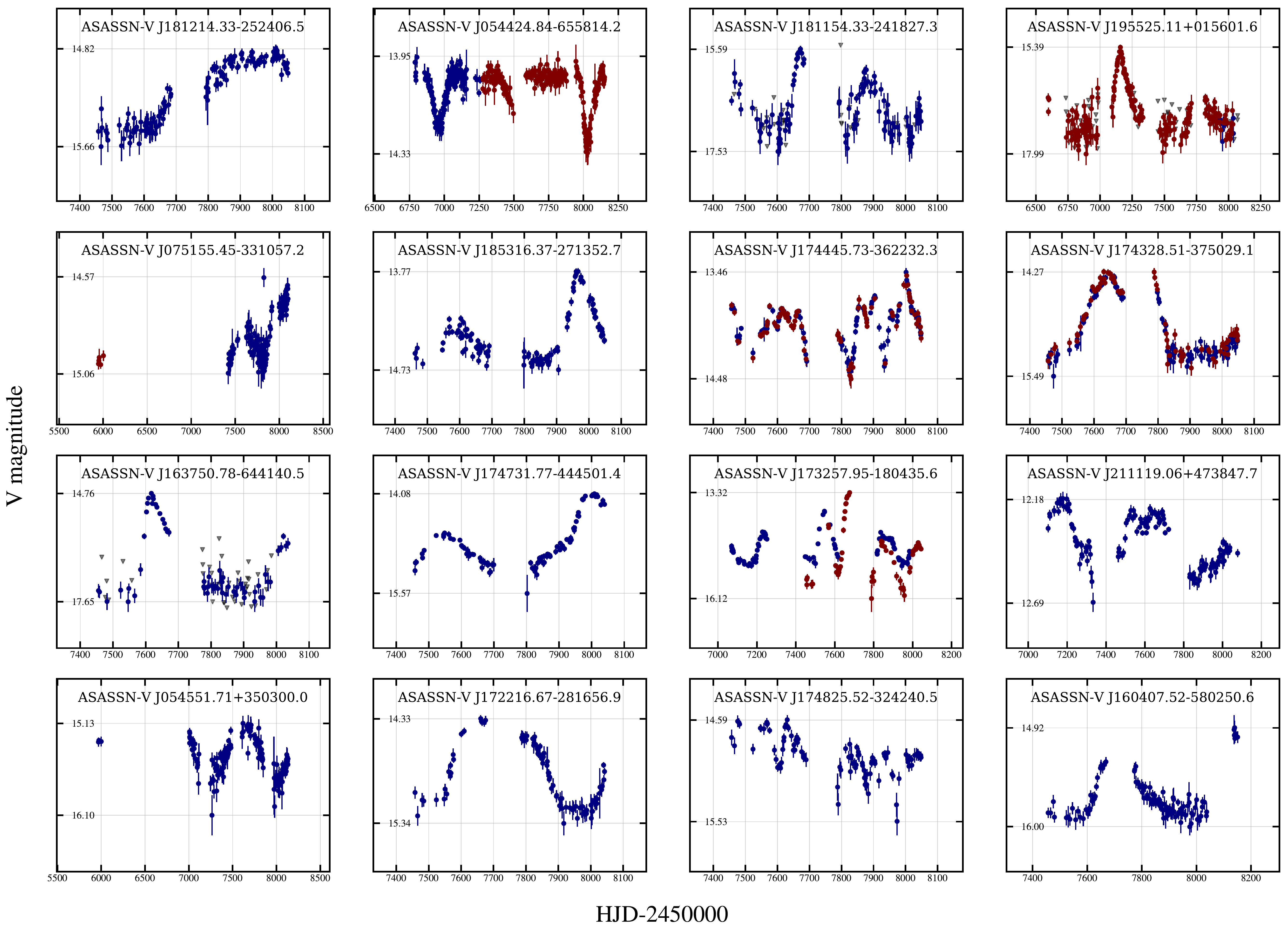}
\caption{ 
	ASAS-SN light curves of variable stars that are weak RCB candidates from Table \ref{table:iffy}. 
  }
\label{fig:rcb_uncertain}
\end{figure*}

\section{Optical Variability}

ASAS-SN has been operating since 2013 (\citealt{2014ApJ...788...48S}) and provides up to 5 years of data to examine for RCB-like variability. It saturates at $V \sim 10 - 11$ mag and can detect objects down to $V \sim 17$ mag (\citealt{2017PASP..129j4502K}). We extracted light curves for all sources in the two  RCB candidates lists from Section \ref{Target Selection}, as well as for all the RCBs reported by SIMBAD (\citealt{2000A&AS..143....9W}).

We began by examining the light curves of known RCBs to understand how they would appear in our data. A typical RCB light curve shows a plateau in brightness that can last for years, before undergoing an abrupt fading event and then slowly recovering to the plateau brightness. DY Per variables usually decline more slowly and have a more symmetrical recovery (e. g., \citealt{2001ApJ...554..298A}).
We then visually scanned each of the 1602 candidates in Tisserand's original list and the 2615 candidates that we generated using the SED matching approach. We discovered the 15 objects presented in Table \ref{table:rcbs} as strong candidates for new RCBs or DY Pers. Some of these objects have preexisting classifications in the International Variable Star Index (VSX, \citealt{2006SASS...25...47W}), and these are noted in Table \ref{table:rcbs}. We present light curves for each of these objects in Figures \ref{fig:rcb1}, \ref{fig:rcb_large}, and \ref{fig:dy_per}. 

In Table \ref{table:old} we present four objects that we discovered while looking through other variables in ASAS-SN data. ASASSN-V J161156.22-575527.2 was included in \citet{Tisserand2012} and was serendipitously discovered in ASAS-SN data by \citet{2017ATel11017....1J} before we began the search described in this work.  Two of the remaining objects display strong RCB-like variability, but were not included in either of our candidate lists because of their colors. Two of these objects have preexisting classifications in VSX. We present the RCB candidate light curves in Figure \ref{fig:rcb_old}. ASASSN-V J175700.51-213934.5 shows DY Per like variability and has been grouped with the other DY Per candidates in Figure \ref{fig:dy_per}. We additionally present 16 more objects with peculiar light curves that are weak RCB candidates. These objects are listed in Table \ref{table:iffy} with speculative variability types based on their light curve morphologies, and their light curves are presented in Figure \ref{fig:rcb_uncertain}.

\section{Discussion}

\begin{figure*}
\centering
\includegraphics[width=\linewidth]{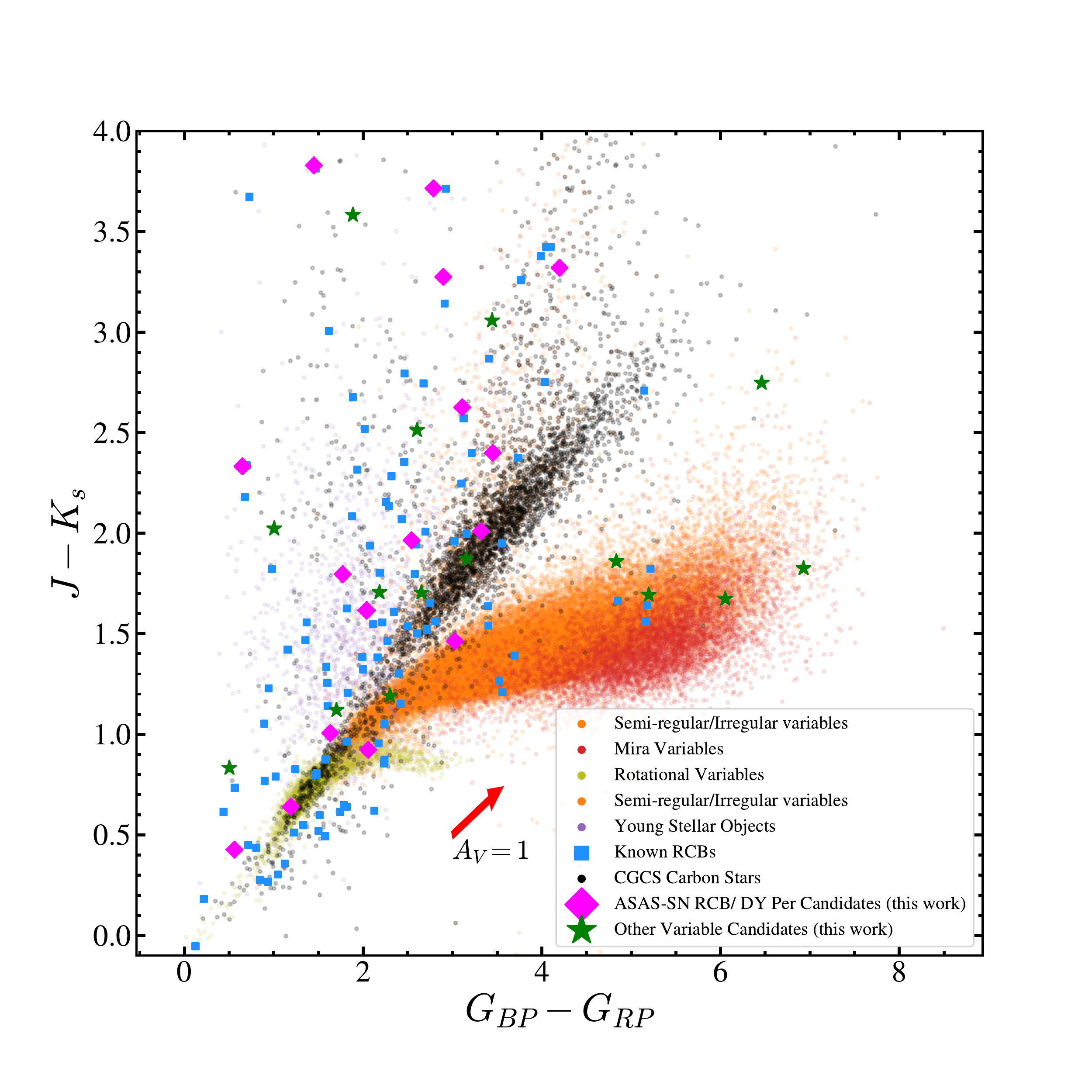}
\caption{ 
Gaia DR2 $G_{BP}$-$G_{RP}$ vs. 2MASS $J-K_s$ color-color diagram. Sources from the Catalog of Galactic Carbon Stars \citep{2001BaltA..10....1A} are colored in black, and sources from the ASAS-SN Catalog of Variable Stars: II \citep{2018arXiv180907329J} are colored by their variability type. RCB candidates from this work are denoted as purple diamonds and peculiar variables in this work are denoted as purple stars. The reddening vector corresponding to an extinction of $A_V=1$ mag is shown in red.
	}
\label{fig:jk}
\end{figure*}

Figure \ref{fig:jk} shows the distribution of RCBs in the Gaia DR2 $G_{BP}-G_{RP}$ vs. $J-K_s$ color-color space \citep{2018arXiv180409365G,2006AJ....131.1163S}. We compare the RCBs with a sample of rotational, Mira, and semi-regular/irregular variables from \cite{2018arXiv180907329J} and the Catalog of Galactic Carbon Stars \citep{2001BaltA..10....1A}. The carbon stars, Mira variables and semi-regular variables all form distinct locii in this color-color space, where the carbon rich sources have redder $J-K_s$ colors for any given $G_{BP}-G_{RP}$ beyond $G_{BP}-G_{RP}\sim2$. RCBs have carbon rich atmospheres and most known RCBs lie on or above the locus of carbon stars in $G_{BP}-G_{RP}$ vs $J-K_s$. A few known RCBs fall along the semi-regular/Mira locii, making these classifications uncertain, although this distinction becomes hazy towards bluer colors. We note that our RCB candidates are consistent with the general distribution of known RCBs.

\begin{figure*}
\centering
\includegraphics[width=\linewidth]{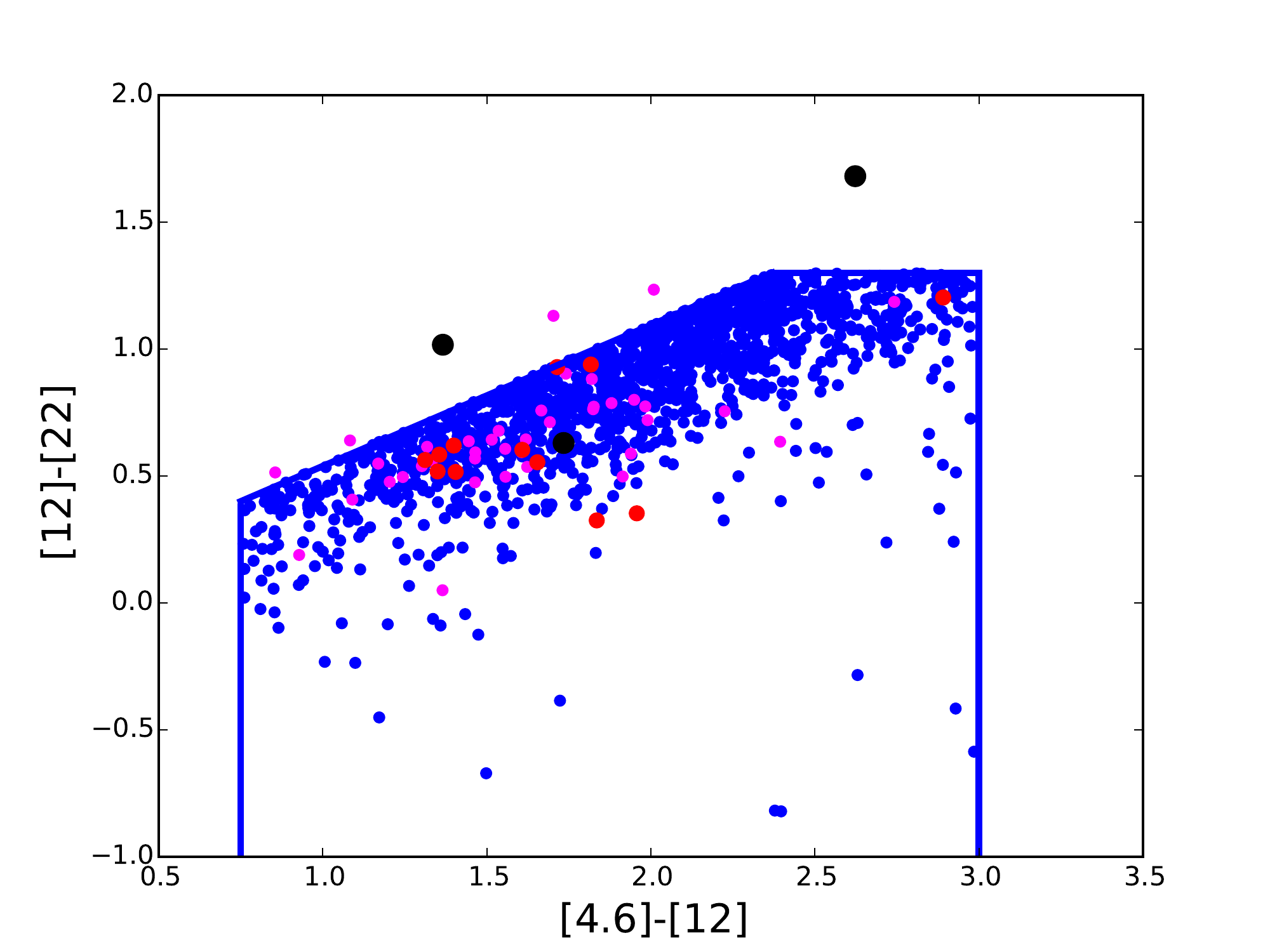}
\caption{ 
[12]-[22] vs. [4.6]-[12] ALLWISE color-color diagram. The blue points are the 1602 candidates from \protect\cite{Tisserand2012}, with the blue lines showing the initial color cuts used to generate the list. The red points are our new RCB candidates, and the black points are the three RCB candidates we discovered outside our initial search. The pink points are previously known RCB candidates from  \protect\cite{Tisserand2012}.  
	}
\label{fig:color}
\end{figure*}

The stars presented in Table \ref{table:old} were discovered outside of our original search. Other than the RCB candidate ASASSN-V J161156.22-575527.2, the remaining RCB candidates have colors that fail the initial mid-IR color cuts used by \cite{Tisserand2012} and our own procedure (see Figure \ref{fig:color}). Their light curves show distinctive RCB like variability making them likely RCB candidates. Their existence suggests that more RCBs should be visually identifiable in ASAS-SN data given a simple method to search for RCB-like variability independent of color information. 

These candidates were selected because they have near/mid-IR spectral energy distributions and optical light curves that are fairly typical of RCBs.  There are other classes of variables which undergo dust formation episodes (\citealt{2014JAVSO..42...13O}) that might be included in the sample, so spectroscopic observations will be necessary for final confirmation of the classifications.

As we were completing this paper, \cite{2018arXiv180901743T} reported the discovery and spectroscopic confirmation of 45 new RCBs. Five of these systems, as well as two of their strong RCB candidates, are on our high confidence list, and one is on the weaker candidate list, as indicated in Tables \ref{table:rcbs} and \ref{table:iffy}. Additionally, \cite{2018arXiv180901474T} updated the infrared selection of \cite{Tisserand2012} that we used in this paper. Our next steps include searching through the light curves of the new candidates.

\section*{Acknowledgments}

We thank the referee, Geoff Clayton, for his comments that helped improve this paper. We thank the Las Cumbres Observatory and its staff for its
continuing support of the ASAS-SN project. We also thank
the Ohio State University College of Arts and Sciences Tech-
nology Services for helping us set up the ASAS-SN variable
stars database.

ASAS-SN is supported by the Gordon and Betty Moore
Foundation through grant GBMF5490 to the Ohio State
University and NSF grant AST-1515927. Development of
ASAS-SN has been supported by NSF grant AST-0908816,
the Mt. Cuba Astronomical Foundation, the Center for Cos-
mology and AstroParticle Physics at the Ohio State Univer-
sity, the Chinese Academy of Sciences South America Cen-
ter for Astronomy (CAS- SACA), the Villum Foundation,
and George Skestos.

This publication makes use of data products from the Two Micron All Sky Survey, which is a joint project of the University of Massachusetts and the Infrared Processing and Analysis Center/California Institute of Technology, funded by the National Aeronautics and Space Administration and the National Science Foundation.
 
This publication makes use of data products from the Wide-field Infrared Survey Explorer, which is a joint project of the University of California, Los Angeles, and the Jet Propulsion Laboratory/California Institute of Technology, funded by the National Aeronautics and Space Administration. 

This research has made use of the NASA/ IPAC Infrared Science Archive, which is operated by the Jet Propulsion Laboratory, California Institute of Technology, under contract with the National Aeronautics and Space Administration.

This work has made use of data from the European Space Agency (ESA)
mission {\it Gaia} (\url{https://www.cosmos.esa.int/gaia}), processed by
the {\it Gaia} Data Processing and Analysis Consortium (DPAC,
\url{https://www.cosmos.esa.int/web/gaia/dpac/consortium}). Funding
for the DPAC has been provided by national institutions, in particular
the institutions participating in the {\it Gaia} Multilateral Agreement.

\clearpage

\begin{table*}
\caption{Candidate ASAS-SN RCB Stars}
\centering
\begin{tabular}{l@{\hskip 0cm} L r@{\hskip 0cm}l R@{\hskip 0cm}L c}
\hline\hline
Name & &  & RA & & \mathrm{Dec} & General Information \\

\hline

ASASSN-V J&053745.71-635330.9^* & 84&.440460 & -63&.891918 & \\
ASASSN-V J&053213.93+340601.4 & 83&.058029 & +34&.100399 & \\
ASASSN-V J&173819.81-203632.2^* & 264&.582550 & -20&.608932 &  \\
ASASSN-V J&173737.08-072828.2 & 264&.404480 & -07&.474488 & \\
ASASSN-V J&174257.20-362052.1^* & 265&.738342 & -36&.347805 & \\
ASASSN-V J&190309.89-302037.0^{**} & 285&.791229 & -30&.343609 &  \\
ASASSN-V J&170737.02-314812.5 & 256&.904236 & -31&.803482 & \\
ASASSN-V J&175031.71-233945.7^* & 267&.632111 & -23&.662706 &  \\
ASASSN-V J&044531.02-683431.3 & 71&.379262 & -68&.575364 & OGLE-LMC-LPV-02510 \\
ASASSN-V J&004822.94-734104.6 & 12&.095596 & -73&.684622 & RAW 476 - Carbon star\\
\hline
\multicolumn{7}{c}{Large Amplitude Candidates}\\

ASASSN-V J&174317.53-182402.5 & 265&.823029 & -18&.400686 & Mira V3062 Oph\\ 
ASASSN-V J&170552.81-163416.6 & 256&.470062 & -16&.571277 & \\
\hline
\multicolumn{7}{c}{DY Per Candidates}\\

ASASSN-V J&191243.07+055313.1^{**} & 288&.179474 & +05&.886981 & \\
ASASSN-V J&175526.28-214214.1 & 268&.859497 & -21&.703928 & \\
ASASSN-V J&202300.80+431111.5 & 305&.753330 & +43&.186518 & \\

\hline

\multicolumn{7}{l}{\textbf{Note} Large amplitude candidates have variability above 3 mag }\\
\multicolumn{7}{l}{$^*:$ confirmed RCB star in \protect\cite{2018arXiv180901743T}}\\
\multicolumn{7}{l}{$^{**}:$ RCB candidate star in \protect\cite{2018arXiv180901743T}}\\

\end{tabular}
\label{table:rcbs}
\end{table*}

\begin{table*}
\caption{RCB Candidates discovered outside of our search}
\centering
\begin{tabular}{l@{\hskip 0cm} L c r@{\hskip 0cm}l R@{\hskip 0cm}L c}
\hline\hline
Name & & VSX Name &  & RA & & \mathrm{Dec} & VSX Classification/Notes \\
\hline
ASASSN-V J&161156.22-575527.2^* & & 242&.98429 & -57&.92422 & previous ASAS-SN discovery\\
\hline
ASASSN-V J&043259.32+415854.0 & HH Per & 68&.24717 & +41&.98168 & long-period variable\\

ASASSN-V J&201504.29+462719.9 & & 303&.76791 & +46&.45553 & \\
\hline
\multicolumn{8}{c}{DY Per Candidate}\\
ASASSN-V J&175700.51-213934.5 & Mis V0832 & 269&.25200 & -21&.65972 & semi-regular variable \\ 
\hline
\multicolumn{8}{l}{$^*:$ confirmed RCB star in \cite{2018arXiv180901743T}}

\end{tabular}
\label{table:old}
\end{table*}

\begin{table*}
\caption{Weak RCB Candidates}
\centering
\begin{tabular}{l@{\hskip 0cm} L r@{\hskip 0cm}l R@{\hskip 0cm}L c}
\hline\hline
Name & &  & RA & & \mathrm{Dec} & Suspected Variability Type \\
\hline

ASASSN-V J&195525.11+015601.6  & 298&.854614 & +1&.933783 & \\
ASASSN-V J&075155.45-331057.2  & 117&.981049 & -33&.182560 & \\
ASASSN-V J&185316.37-271352.7  & 283&.318207 & -27&.231319 & \\
ASASSN-V J&174445.73-362232.3  & 266&.190521 & -36&.375629 & semi-regular variable\\
ASASSN-V J&174328.51-375029.1^*  & 265&.868774 & -37&.841412 & long-period variable or RCB\\
ASASSN-V J&163750.78-644140.5  & 249&.461594 & -64&.694595 & \\ 
ASASSN-V J&174731.77-444501.4  & 266&.882360 & -44&.750380 & \\
ASASSN-V J&173257.95-180435.6  & 263&.241443 & -18&.076560 & semi-regular variable\\
ASASSN-V J&211119.06+473847.7  & 317&.829426 & +47&.646576 & irregular variable\\
ASASSN-V J&054551.71+350300.0  & 86&.465439 & +35&.050001 & irregular variable\\
ASASSN-V J&172216.67-281656.9  & 260&.569464 & -28&.282466 & long-period variable\\
ASASSN-V J&174825.52-324240.5  & 267&.106334 & -32&.711251 & irregular variable\\
ASASSN-V J&160407.52-580250.6  & 241&.031323 & -58&.047382 & \\

ASASSN-V J&181154.33-241827.3 & 272&.976368 & -24&.307591 & semi-regular variable\\

ASASSN-V J&054424.84-655814.2 & 86&.103507 & -65&.970622 & \\ 

ASASSN-V J&181214.33-252406.5 & 273&.059723 & -25&.401814 & \\

\hline

\multicolumn{7}{l}{$^*:$ confirmed RCB star in \protect\cite{2018arXiv180901743T}}
\end{tabular}
\label{table:iffy}
\end{table*}

\end{document}